\documentclass[twocolumn,showpacs,aps,pre]{revtex4-1}

\usepackage{ifthen}
\usepackage{ifpdf}
\usepackage{color}

\ifpdf
\usepackage{graphicx}
\usepackage{epstopdf}
\else
\usepackage{graphicx}
\usepackage{epsfig}
\fi
\graphicspath{{./Figs/}{./}}

\usepackage{latexsym}
\usepackage{amsmath}
\usepackage{amstext}
\usepackage{amssymb}
\usepackage{bm}
\usepackage{wasysym}
\usepackage[pdftitle={},bookmarks]{hyperref}

\newcommand{\amatrix}[1]{\begin{matrix} #1 \end{matrix}} 
\newcommand{\eexp}[1]{\mathrm{e}^{#1}}

\newcommand{\braket}[1]{ \left\langle #1 \right\rangle}

\newcommand{\be}[1]{\begin{eqnarray} \label{e#1}}
\newcommand{\beq}{\begin{eqnarray}}
\newcommand{\eeq}{\end{eqnarray}} 

\newcommand{\hide}[1]{}

\newcommand{\Eq}[1]{{\textcolor{blue}{Eq.}}(\ref{#1})} 
\newcommand{\Fig}[1] {{\textcolor{blue}{Fig.}}\ref{#1}}

\newcommand{\hrefl}[1]{\href{#1}{[link]}}

\begin{document}

\title{Resistor-network anomalies in the heat transport of random harmonic chain}

\author{Isaac Weinberg$^1$, Yaron de Leeuw$^1$, Tsampikos Kottos$^{2,3}$, Doron Cohen$^1$}
\affiliation{$^1$Ben-Gurion University of the Negev, Beer-Sheva 84105, Israel}
\affiliation{$^2$Department of Physics, Wesleyan University, Middletown, Connecticut 06459}
\affiliation{$^3$Department of Mathematics, Wesleyan University, Middletown, Connecticut 06459}

\begin{abstract}
We consider thermal transport in low-dimensional disordered harmonic networks of coupled masses. 
Utilizing known results regarding Anderson localization, we derive 
the actual dependence of the thermal conductance~$G$ on the length $L$ of the sample.  
This is required by nanotechnology implementations because for such networks 
Fourier's law ${G \propto 1/L^{\alpha}}$ with ${\alpha=1}$ is violated.  
In particular we consider ``glassy" disorder in the coupling constants, 
and find an anomaly which is related by duality to the Lifshitz-tail regime
in the standard Anderson model. 
\end{abstract}


\maketitle

\section{Introduction}

The theory of phononic heat conduction in disordered low-dimensional networks is a central theme of research in recent years \cite{LLP03,
D08,LRWZHL12}. The interest in this theme is not only purely academic, but it is also motivated by the ongoing developments in nanotechnology.
In spite of the recent research efforts, the understanding of thermal transport is still at its infancy. This becomes more obvious if one compares 
with the achievements that have been experienced during the last fifty years in understanding and managing electron transport. In this respect 
even the microscopic laws that govern heat conduction in low dimensional systems have only recently start being scrutinized via both theoretical, 
numerical and experimental studies \cite{LLP03,D08,COGMZ08,NGPB09,LRWZHL12,ZL10,K1,K2}. These studies unveil many surprising results, the most 
dramatic of which is the violation of the naive expectation (Fourier's law) which states that the thermal conductance $G$ is inverse proportional 
to the size~$L$ of the system, namely, $G\propto 1/L^{\alpha}$ with ${\alpha=1}$. 

Currently it is well established that in low-dimensional disordered systems, in the absence of non-linearity, Fourier's law  is violated. The
underlying physics is related to the theory of Anderson localization of the vibrational modes \cite{D08,D01,LXXZL12,DL08,LD05,RD08,LZH01,
KCRDLS10a,KCRDLS10b}. On the basis of the prevailing theory \cite{D08,D01} it has been claimed that for samples with ``optimal" contacts ${\alpha=1/2}$, 
while in general $\alpha$ might be larger, say ${\alpha=3/2}$ for samples with ``fixed boundary conditions". Recently the ``optimal" value 
${\alpha=1/2}$ has been challenged by the numerical study of \cite{BZFK13}. These authors found a super-optimal value ${\alpha \sim 1/4}$ for 
moderate system sizes $L$, while asymptotically, in the presence of a pinning potential, $G$ decays exponentially as ${\exp(-\gamma L)}$.

It is obvious that if the final goal 
is to achieve the control of heat flow on the nanoscale, 
first we have to understand the fundamental mechanisms of heat conduction, 
and provide an adequate description of its scaling with the system size for any~$L$, 
including the experimentally relevant cases of intermediate lengths.

\section{Scope}

Considering heat transport for low-dimensional disordered networks of coupled harmonic masses, we utilize known results from the field of mesoscopic electronic physics, in order to derive the actual $L$~dependence of $G$ for regular as well as for ``glassy" type of disorder. The information about the latter is encoded in the dependence of the inverse localization length $\gamma$ on the vibration frequency $\omega$. Our results explain the transition from optimal to super-optimal scaling behavior and eventually to exponential dependence on $L$. We address the implications of the percolation threshold, and the geometrical bandwidth. Along the way we highlight a surprising anomaly that is related by duality to the Lifshitz-tail regime in the standard Anderson model, and test the borders of the one-parameter scaling hypothesis.  

The outline of this paper is as follows:
Sections III-V define the general model of interest,  
emphasizing that for ``glassy disorder" a resistor-network 
perspective is essential. 
Section VI clarifies that the analysis of heat conduction 
of quasi one dimensional networks 
effectively reduces to the analysis of a single-channel problem. 
Section VII explains how we use the transfer matrix method 
in the numerical analysis: we highlight the procedure  
for the determination of the optimal leads, and the
significance of the percolation parameter~$s$ in this context.     
Section VIII use the Born approximation in order 
to provide an explanation for the numerical findings 
of \cite{BZFK13}. These results had been obtained    
for weak disorder.

Subsequently we focus on the single-channel model.   
Our main interest is to explore the implications 
of ``glassy" disorder, and to highlight the resistor-network aspect. 
In Sections IX and X we go beyond the born approximation
by establishing a duality between glassy off-diagonal disorder
and weak diagonal disorder. 
Consequently we deduce that the Lifshitz-tail anomaly
is reflected in the frequency~dependence of the
inverse localization length. 
This prediction is verified numerically.

The remaining sections XI to XIII clarify how scaling-theory 
of localization can be used in order to calculate the heat conductance. 
Here no further surprises are found. In fact we verify numerically 
that a straightforward application of the weak-disorder analytical 
approach is quite satisfactory. In spite of the ``glassy" disorder 
the deviations from one-parameter scaling are not alarming.

\section{The model}

We consider a one-dimensional network of $L$ harmonic oscillators of equal masses. 
The system is described by the Hamiltonian 
\begin{equation}
{\cal H} = {1\over 2} P^T P + {1\over 2} Q^T \bm{W} Q 
\label{Hmatrix} 
\end{equation}
where $Q^T\equiv(q_1,q_2,\cdots,q_N)$, and $P^T\equiv(p_1,p_2,\cdots,p_N)$ 
are the displacement coordinates and the conjugate momenta. 
The real symmetric matrix  $\bm{W}$ is determined by the spring constants.
Its off-diagonal elements $W_{nm}{=}-w_{nm}$ originate from 
the coupling potential $(1/2)\sum_{m,n}w_{nm}(q_n-q_m)^2$, 
while its diagonal elements contain an additional 
optional term that originate from a pinning potential ${(1/2)\sum_n v_n q_n^2}$ 
that couples the masses to the substrate. 
Accordingly ${W_{nn}=v_n+\sum_m w_{nm}}$. 
For a chain with near-neighbor transitions we use 
the simplified notation ${w_{n{+}1,n} \equiv w_n}$.

In general the interest is in quasi one-dimensional networks, 
for which $\bm{W}$ is a banded matrix with ${1{+}2b}$ diagonals.
For ${b{=}1}$ the near-neighbor hopping implies a single-channel system. 
For ${b>1}$ the dispersion relation (see section~VI below) 
has several branches, which is like having a multi-channel system.
The heat conduction of such networks has been investigated numerically 
in \cite{BZFK13}, with puzzling findings that have not been explained 
theoretically. We shall see that the essential physics can be 
reduced to single channel ($b{=}1$) analysis. On top we would like  
to consider not only weakly disordered network, but also 
the implications of ``glassy" disorder as defined below.

\section{The disorder}

Both the $w_{nm}$ and the $v_n$ are assumed to 
be random variables. The diagonal-disorder due to the pinning
potential is formally like that of the standard Anderson model 
with some variance ${\sigma_{\parallel}^2 \equiv \mbox{Var}(v)}$.    
The off-diagonal disorder of the couplings might be 
weak with some variance ${\sigma_{\perp}^2 = \mbox{Var}(w)}$, 
but more generally it can reflect the glassiness of the network. 
By ``glassy disorder" we mean that the coupling~$w$ has an 
exponential sensitivity to physical parameters.  
For {\em random barrier} statistics ${ w \propto \eexp{-B} }$, 
where $B$ is uniformly distributed within $[0,\sigma]$, accordingly 
\be{3002}
P(w) \ \ \propto \ \ \frac{1}{w} \ \ \left(\eexp{-\sigma} < \frac{w}{w_c} <1\right)
\eeq
For {\em random distance} statistics ${ w \propto \eexp{-R} }$,
where $R$ is implied by Poisson statistics. 
The probability distribution in the latter case is 
\be{3003}
P(w) \ \ = \ \ \frac{s}{w_c^{s}} w^{s-1} \ (w<w_c)
\eeq
where~$s$ is the normalized density of the sites.
Large $s$ is like regular weak disorder, while small $s$ 
implies glassy disorder that features log-wide distribution  
(couplings distributed over several orders of magnitude). 
The case $s=0$ with an added lower cutoff 
formally corresponds to ``random barriers".

\section{Resistor-network perspective}

It is useful to notice that the problem of phononic heat conduction 
in the absence of a pinning potential is formally equivalent 
to the analysis of a rate equation, where the spring-constants 
are interpreted as the rates~$w_{nm}$ 
for transitions between sites~$n$ and~$m$. 
Optionally it can be regarded as a resistor-network problem 
where $w_{nm}$ represent connectors. 
We define $w_0$ as the effective hopping rate between sites. 
We later justify that it should be formally identified with  
the {\em conductivity} of the corresponding resistor-network. 

The detailed numerical analysis in the subsequent
sections concerns the ${b=1}$ chain, 
for which the ``serial addition" rule implies 
that $w_0$ equals the harmonic average.
For the ``random distance" disorder of \Eq{e3003} we get 
\beq
w_0 \ = \ \left[ \braket{\frac{1}{w}} \right]^{-1} \ = \ \ (s{>}1)\left[\frac{s-1}{s}\right] w_c
\eeq
For ${s<1}$ the network is no longer percolating, namely ${w_0=0}$.  
In the present context $w_0$ determines the speed of sound (see below).

For the later analysis we need also the second moment
of the couplings. For $s>2$ one obtains 
\be{633}
\braket{\left(\frac{1}{w}\right)^2}_{s{>}2} \ = \ \left[\frac{s}{s-2}\right] \left(\frac{1}{w_c}\right)^2 
\eeq 
Hence the variance is $[s/((s{-}1)^2(s{-}2))]w_c^{-2}$. 
For ${s<2}$ the second moment diverges. 
But for a particular realization the sample-specific  
result is finite, and depends on the effective lower 
cutoff $\delta w$ of the distribution.
The number ${\mathcal{N}\equiv w_c/\delta w}$      
reflects the finite size of the sample, and we get
the sample-size dependent result 
\be{634}
\braket{\left(\frac{1}{w}\right)^2}_{s{<}2} \ = \ \frac{s}{2{-}s}\left[\mathcal{N}^{2{-}s}-1\right] \left(\frac{1}{w_c}\right)^2 
\eeq 
As we go from ${s>2}$ to ${s<2}$ the dependence 
of the variance on $s$ has a crossover from power-law to exponential. 
We shall see later that this crossover is reflected 
in the localization-length of the eigenstates.

\begin{figure*}
\centering
\includegraphics[width=0.9\hsize]{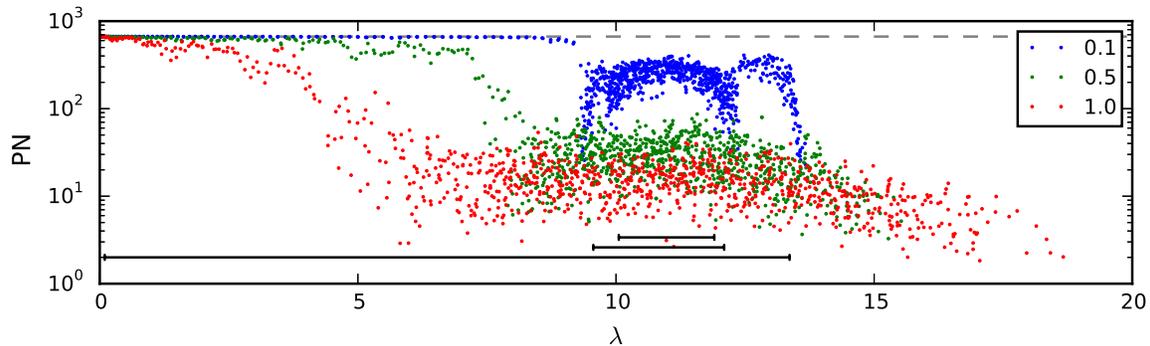}
 
\caption{
The participation number (PN) \cite{rmrkA} of the eigenstates of a conservative banded matrix 
are plotted against their eigenvalues $\lambda$.
The $0< |m-n| \le b$ elements of $\bm{W}$ are random numbers $w\in[1-\varsigma,1+\varsigma]$ (box distribution).
The values of $\varsigma$ are indicated in the legend  
(note that the numerics of \cite{BZFK13} corresponds to $\varsigma{=}0.5$). 
The number of bands is $b{=}5$, and the length of the sample is $N{=}1000$ with periodic boundary conditions. 
The support of the clean-ring channels is indicted by the lower horizontal lines.
We observe the gradual blurring of the non-disordered band structure. 
All the eigenstates that have large PN reside in the lower part of the spectrum and belong to a single channel. 
}

\label{f1}
\end{figure*}

\section{The spectrum}

The eigenvalues $\lambda_k$ are determined via diagonalization $\bm{W}Q=\lambda Q$, 
from which one deduces the eigen-frequencies via ${\lambda_k=\omega_k^2}$.
In the absence of disorder the eigenmodes are Bloch states with 
\be{3005}
\lambda_k \ = \ 2w_0 \sum_{r=1}^{b}[1-\cos(r k)] \ \ \equiv \ \ \omega_k^2
\eeq  
where~$k$ is the associated wavenumber. 
For a single-channel $\lambda=2w_0(1{-}\cos k) \approx w_0 k^2$, 
where the small-$k$ approximation holds close to the band floor.
With disordered couplings, but in the absence of a pinning potential 
the lowest eigenvalue is still ${\lambda_0=0}$,  
which corresponds to the trivial extended state ${Q=(1,1,...,1)}^T$, 
that is interpreted as the ergodic state in the context of rate equations. 
All higher eigenstates are exponentially localized, and are characterized 
by a spectral density $\varrho(\omega)$.

In \Fig{f1} we provide a numerical example considering a ${b=5}$ quasi-one dimensional sample. 
The dispersion relation \Eq{e3005} has 5~branches. 
The support of the 1st, 3rd and 5th ascending branches is indicated in the figure. 
{\em It is important to observe that at the bottom of the band 
a single channel-approximation is most appropriate.}
Hence within the framework of the Debye approximation 
the dispersion at the bottom of the band is always 
\beq
\omega \ \ \approx  \ \ c k
\ \ \ \ \ \ \ \text{[Debye]}
\eeq
For $b{=}1$ the speed of sound is ${c=\sqrt{w_0}}$, 
while for $b{\gg1}$ it is easily found that 
\beq
c \ \ \approx \ \ [(1/3)b^3]^{1/2} \, \sqrt{w_0}
\eeq
Either way the low-frequency spectral density is constant, namely 
\beq
\rho(\omega) \ \ \approx \ \  \frac{L}{\pi c}
\eeq
The effect of weak disorder on this result is negligible.

\section{Localization}

\begin{figure}
\centering
\includegraphics[width=\hsize]{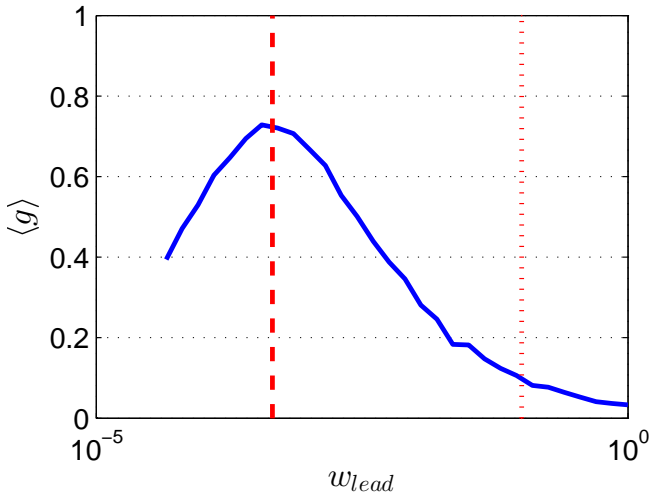}
 
\caption{
The average transmision $\braket{g}$ as a function of the hopping rate $w_{\text{lead}}$ within the leads.
The calculation is done for ${L=50}$ disordered sample,
where the $w_n$ are distributed according to \Eq{e3002} with ${\sigma=10}$ at ${k=0.028\pi}$. 
The algebraic and the harmonic mean values of the~$w_n$ are indicated 
by vertical dotted and dashed lines respectively.}

\label{f15}
\end{figure}

The disorder significantly affects the eigenmodes: rather than being extended 
as assumed by Debye, they become exponentially localized. 
We use the standard notation $\gamma(\omega)$ for the inverse localization length.
Considering a single-channel (${b{=}1}$) system it is defined 
via the asymptotic dependence of the transmission~$g$ on the length~$L$ of the sample.
Namely,    
\beq
\label{locdef}
\gamma(\omega) \ \ = \ \ -\lim_{L\rightarrow \infty}{1\over 2}{\langle \ln(g) \rangle_{\omega} \over L}
\eeq
where $\langle\cdots\rangle$ indicates an averaging over disorder realizations. 
The notion of transmission is physically appealing here,  
because we can regard~$\bm{W}$ as the Hamiltonian of an electron in a tight binding model. 
The transmission can be calculated from the transfer matrix $\bm{T}$ of the sample:
\beq
\label{gTMM}
g = \frac{4|\sin(k)|^2}{|T_{21}-T_{12}+T_{22}\exp(ik)-T_{11}\exp(-ik)|^2}
\eeq
where 
\be{5}
\bm{T} \ \ = \ \ \prod_{n=1}^{n=L} 
\left(\amatrix{
\frac{\lambda-(v_n + w_{n}+w_{n+1})}{w_{n+1}} & -\frac{w_{n}}{w_{n+1}} \cr 1 & 0 
}\right)
\eeq
Above it is assumed that the sample is attached to two non-disordered leads.
Optimal coupling requires the hopping-rates there to be all equal 
to the ``conductivity"~$w_0$, meaning same speed of sound. This observation 
has been verified numerically, see \Fig{f15}. 
We see that it is the resistor-network harmonic-average 
and not the algebraic-average that determines the optimal coupling.

\begin{figure}
\centering
\includegraphics[width=\hsize]{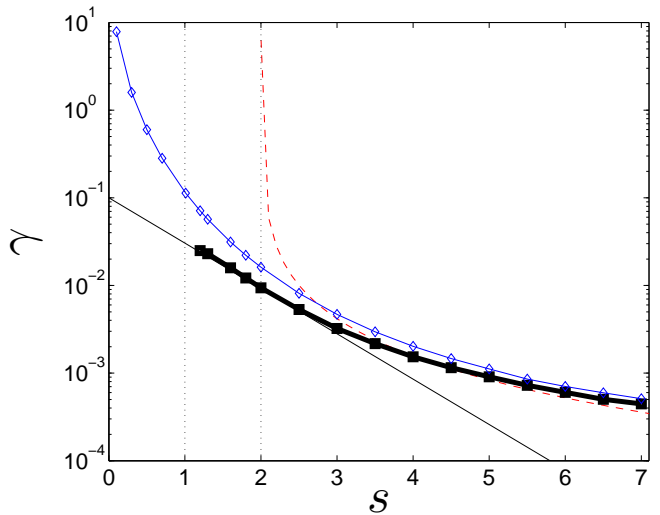}
 
\caption{
The inverse localization length $\gamma$ 
for the {\em random distance} disorder of \Eq{e3003} versus $s$.
Solid line with diamonds is for the non-optimized (${w_{\text{lead}}=w_c}$) results. 
Thick solid line with squares is for the optimized (${w_{\text{lead}}=w_0}$) results in the range~${s>1}$, 
where $w_0$ is finite.  
The tangent thin solid line has been fitted in the range ${1<s<2}$ where \Eq{e634} is expected to hold.
The naive Born approximation \Eq{FGRw} is illustrated by dashed line in the range ${s>2}$.
The numerical results here and in the next figures 
are based on the transfer matrix method (symbols), 
with several hundreds of realizations up to ${L\sim10^4}$.}
\label{f16}
\end{figure}

In \Fig{f16} we display an example for the calculation of $\gamma$ versus~$s$. 
Well-defined results are obtained for ${s>2}$ where the second moment \Eq{e633} is finite. 
In the next section we shall derive a naive Born approximation for $\gamma$. 
This is displayed in \Fig{f16} too as a dashed line. 
The estimate is based on analytical ensemble-average of $\text{Var}(1/w)$ 
and therefore diverges as ${s=2}$ is approached from above.
In the range ${1<s<2}$ the second moment \Eq{e634} is ill-defined (sample-specific).
Given an individual sample the Born approximation can be used 
with sample-variance (which is always finite) and provide a rough estimate. 
The typical result in this range is expected to depend exponentially on~$s$ 
as implied by \Eq{e634}. This expected dependence is indeed observed.
For ${s<1}$ the $s$~dependence of $\gamma$ is completely ill-defined: the chain is non-percolating 
in the $L\to\infty$ limit, and the contact optimization procedures becomes meaningless.

\section{Born approximation}

In the absence of disorder $\bm{W}$ describes hopping 
with some rate $w_0$, and the eigenstates are free waves labeled by~$k$. 
With disorder the~$w_0$ of the unperturbed Hamiltonian is 
loosely defined as the average~$w$. Later we shall go beyond 
the Born approximation and will show that it should 
be the harmonic average (as already defined previously).   
The disorder couples states that have different~$k$. 
For diagonal disorder ("pinning") the couplings are proportional to the 
variance of the diagonal elements, 
namely ${ \overline{|W_{k,k'}|^2} = (1/L)\text{Var}(v)}$.
For off-diagonal disorder (random spring constants) 
the  couplings are proportional to the variance  of 
the off-diagonal elements, and depends on~$b$ and on~$k$ too:
\beq \nonumber
\overline{\left|W_{k',k}\right|^2}  =  \frac{\text{Var}(w)}{L} \ \sum_{r=1}^{b} \left[2\sin(rk'/2) \, 2\sin(rk/2)\right]^2 
\eeq
It follow that for small~$k$ we have ${|W_{k,k'}|^2 \propto b^5 \sigma_{\perp}^2 k^4}$.

\begin{figure}

\includegraphics[width=0.85\hsize]{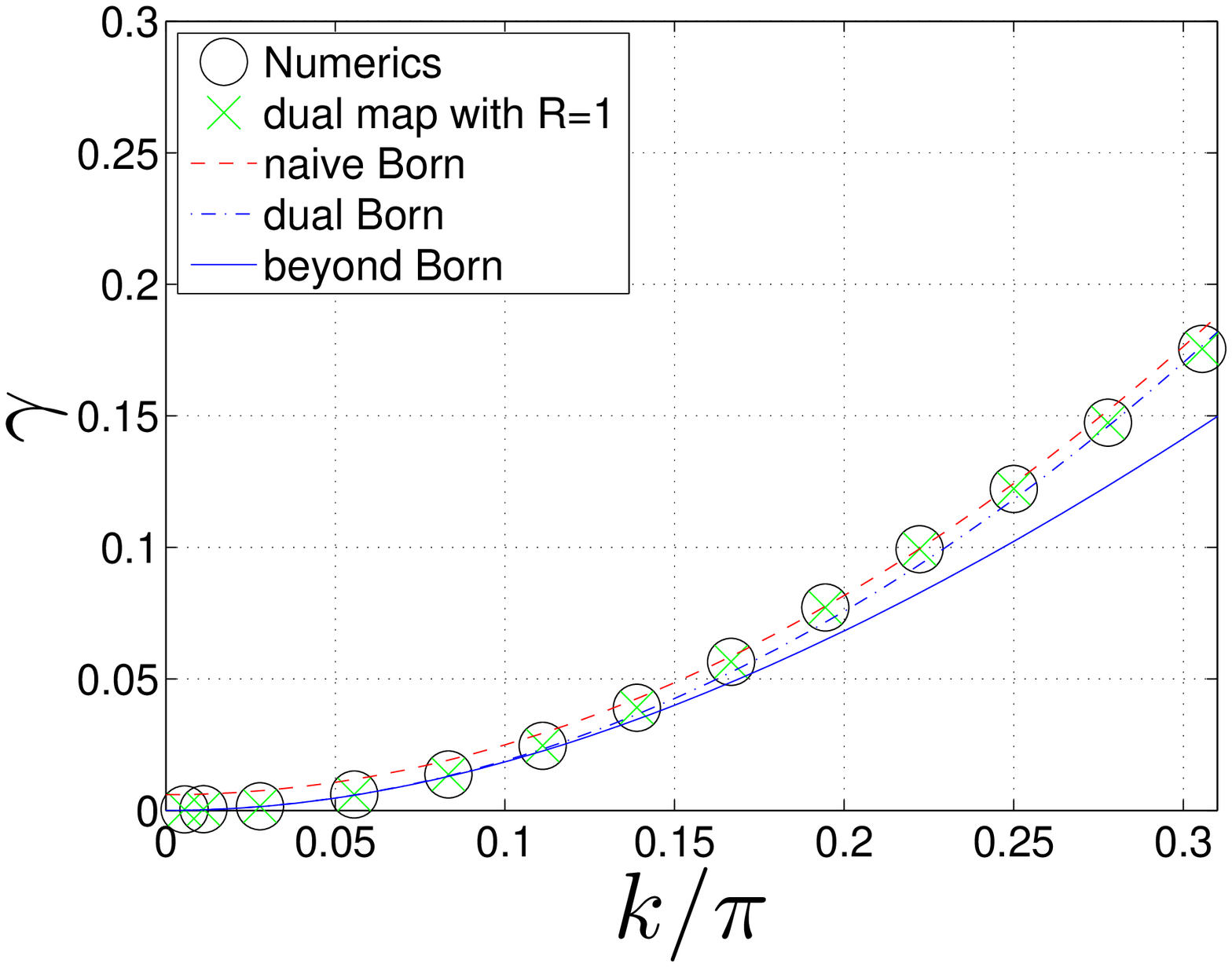} 

\includegraphics[width=0.85\hsize]{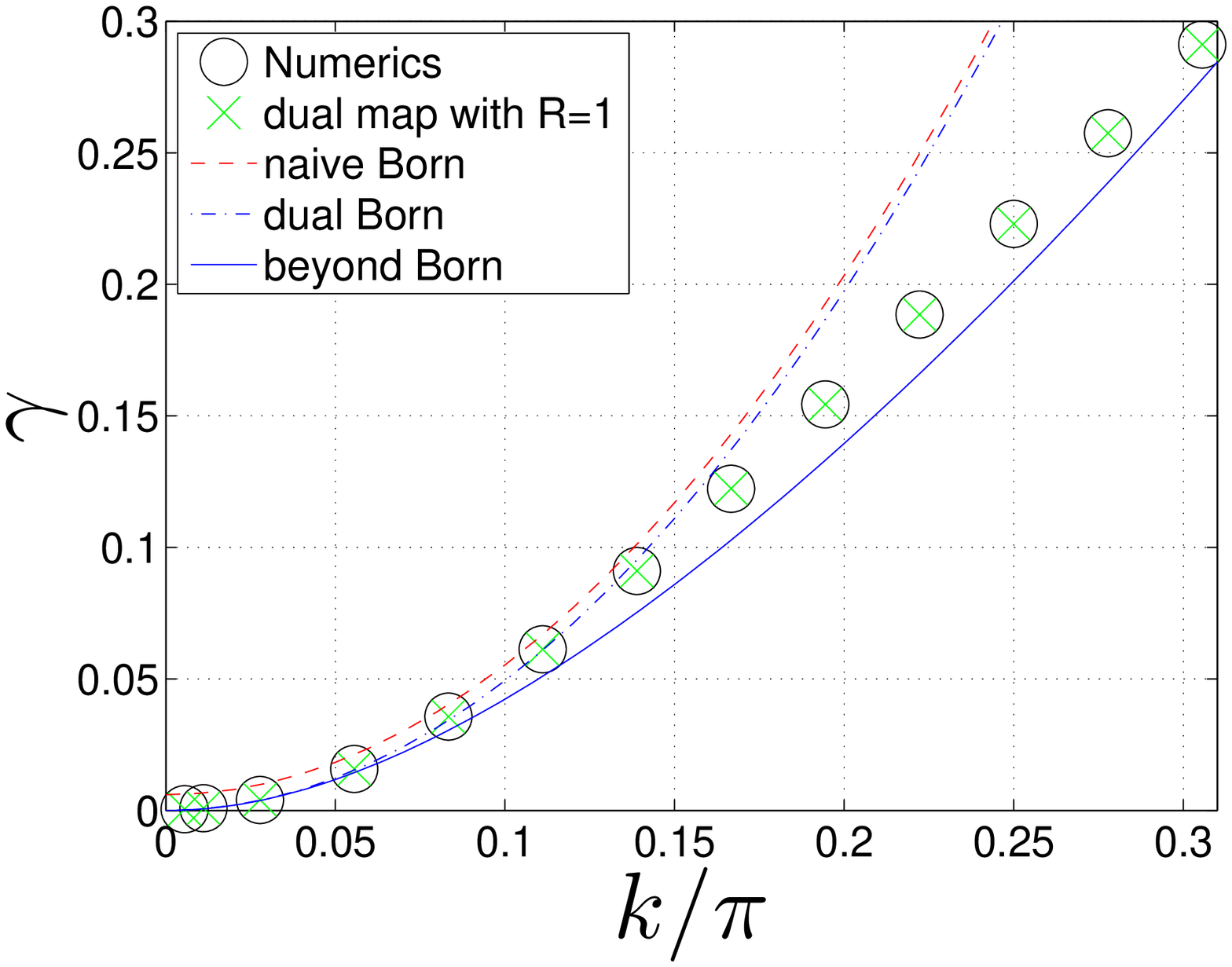}

\includegraphics[width=0.85\hsize]{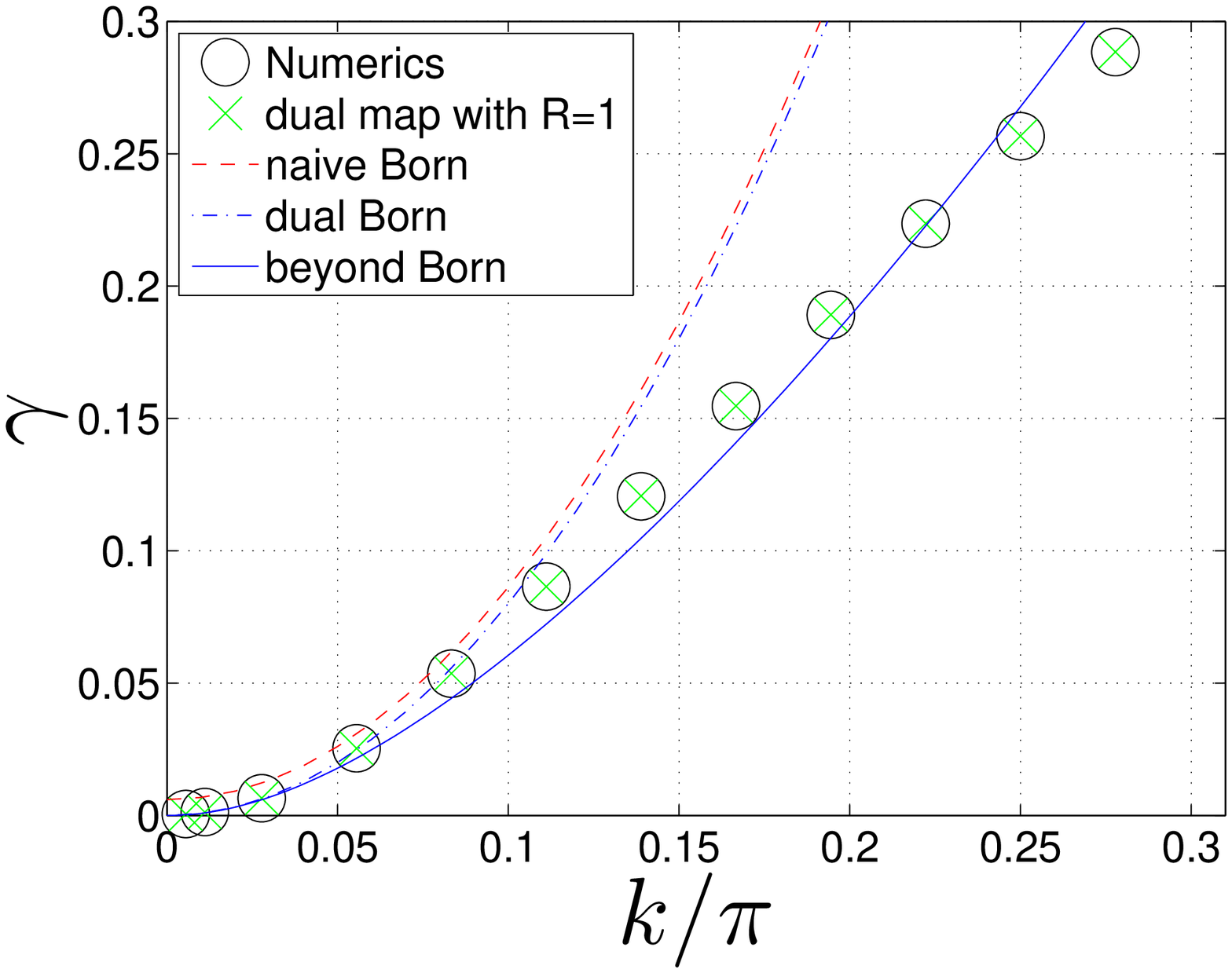}

\caption{
The inverse localization length~$\gamma$ versus~$k$ 
for the {\em random barrier} disorder of \Eq{e3002} 
with $\sigma=5,10,15$ from up to down.    
Pure off-diagonal disorder is assumed. 
The numerical results based on the transfer matrix method (circles)
are compared with those that are generated by the map \Eq{e10} 
with the approximation ${R_n{=}1}$.
The naive Born estimate \Eq{FGRw} is illustrated by dashed red line, 
while the blue dashed-dotted line is based on the improved estimate with \Eq{e12}.
Here we consider the distribution of \Eq{e3002} for which both estimates coincide identically, 
therefore a small shift has been inserted artificially so the two lines could be discern.
The inverse localization length $\gamma$ is over-estimated 
as $k$ becomes larger due to the Lifshitz tail anomaly.
The solid blue line is based on \Eq{e151} with no fitting parameters.    
}
\label{f2a}
\end{figure}

\begin{figure}

\includegraphics[width=0.85\hsize]{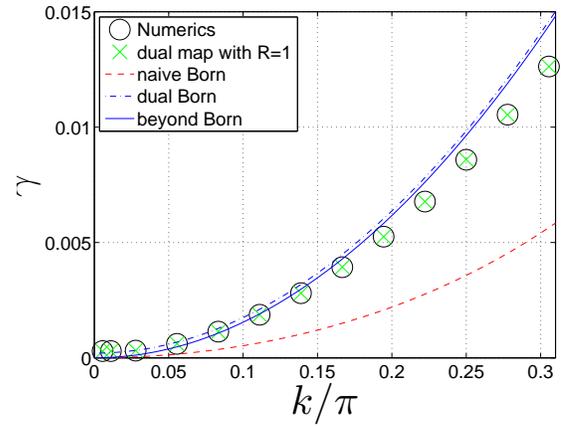}

\caption{
The inverse localization length $\gamma$ versus $k$ 
for the {\em random distance} disorder of \Eq{e3003} with $s=4$.      
The symbols and lines are the same as in \Fig{f2a}.
Note that the ``dual Born" approximation  
cannot be resolved from its improved version \Eq{e151}.
Here we consider the distribution of \Eq{e3003} for which 
the duality implies $k$~dependence that {\em does not coincide} 
with that of the naive Born approximation.
}
\label{f2b}
\end{figure}

The Fermi-Golden-Rule (FGR) picture implies that 
the scattering rate is ${\tau^{-1}= 2\pi \varrho(\omega) \overline{\left|W_{k',k}\right|^2}}$. 
The Born approximation for the mean free path is ${\ell=[d\lambda/dk]\tau}$, 
where the expression in the square brackets is the group velocity 
in the electronic sense ($\lambda$ is like energy). 
The Debye approximation implies $d\lambda/dk \approx 2[c^2]k$.
The inverse localization length is $\gamma=(2\ell)^{-1}$. 
From here (without taking the small $k$ approximation) it follows that 
\beq \label{FGRv}
\gamma(\omega) \ \ &\approx& \ \ 
\frac{1}{8}\left[\frac{9}{b^6}\right]\left(\frac{\sigma_{\parallel}}{w_0}\right)^2  \left(\frac{1}{\sin(k)}\right)^2
\\ \label{FGRw} &+& \ \ 
\frac{1}{8}\left[\frac{9}{5b}\right]\left(\frac{\sigma_{\perp}}{w_0}\right)^2  \left( 2\tan\left(\frac{k}{2}\right) \right)^2
\eeq
where the prefactors in the square brackets assume ${b\gg1}$, and should be replaced by unity for ${b=1}$.
In the absence of pinning the localization length diverges ($\gamma \propto k^2$) 
at the band floor, as assumed by Debye. This behavior is demonstrated in \Fig{f2a} and \Fig{f2b}
for two types of glassy disorder. The deviations from \Eq{FGRw} 
will be explained in the next paragraphs.

\section{Beyond Born}

The Born approximation has assumed weak disorder. 
Here we would like to consider the more general case 
of glassy disorder. For this purpose, as in~\cite{DG84}, 
we write the equation ${\bm{W}\psi = \lambda \psi}$ for the eigenstates 
as a map of a single variable ${r_n=\psi_{n}/\psi_{n-1}}$,   
namely  ${r_{n+1} = -R_n/r_n - A_n}$,  
where ${R_n=w_{n-1}/w_{n}}$, and ${A_n = (\lambda-v_n-w_n-w_{n-1})/w_n}$.
In the case of diagonal disorder it takes the from  
\be{8}
r_{n+1} \ = \ -\frac{1}{r_n} - \epsilon + f_n
\eeq
where $f_n=v_n/w_0$ is the scaled disorder and 
\be{81}
\epsilon \ \ = \ \ \frac{\lambda}{w_0} - 2 \ \ \equiv \ \  -2\cos(k)  
\eeq
is the scaled energy measured from the center of the band.
Without the random term $f_n$ this map has a fixed-point 
that is determined by the equation ${ r^2+\epsilon r +1=0 }$, 
with elliptic solution for ${\epsilon\in[-2,2]}$. 
The random term is responsible for having 
a non zero inverse localization length ${\gamma = -\langle \ln(r) \rangle}$. 
The Born approximation \Eq{FGRv} is written as  
\be{9}
\gamma \ \ \approx \ \ \frac{1}{8} \, \frac{ \mbox{Var}(f)}{\left[1-(\epsilon/2)^2\right]}
\ \ = \ \ \frac{1}{8} \, \frac{ \mbox{Var}(f)}{ [\sin(k)]^2}
\eeq
This standard estimate does not hold close to the band-edge $\epsilon_0{=}{-}2$,
which can be regarded as an anomaly~\cite{DG84}. 
Closeness to the band-edge means that~${|\epsilon-\epsilon_0|}$ 
becomes comparable with the kinetic energy~$\gamma^2$. 
Hence the so called Lifshitz tail region is  ${|\epsilon-\epsilon_0|<\epsilon_c}$ 
with $\epsilon_c = [\mbox{Var}(f)]^{2/3}$.
Optionally this energy scale can be deduced by dimensional analysis.  
In the Lifshitz tail region the inverse localization 
length has finite value $\gamma \sim \sqrt{\epsilon_c}$.
An analytical expression can be derived using white-noise 
approximation (see Appendix~A for details):  
\be{150}
\gamma \ = \ \left( \frac{1}{2} \text{Var}(f) \right)^{\frac{1}{3}} 
\mathcal{K} \left[ \Big( 2\text{Var}(f)^2 \Big)^{-\frac{1}{3}} \, k^2 \right]
\ \ \ \ \ \ \ \ \ 
\eeq
Outside of the Lifshitz tail region 
this expression reduces back to \Eq{e9}.
To be more precise, if we want to take 
the exact dispersion into account 
an add-hock improvement of \Eq{e150} would be 
to  to replace $k^2$ by $[\sin(k)]^2$. 
But our interest is in small $k$~values, 
for which this improvement is not required in practice:
this has been confirmed numerically (not displayed).

\section{Duality}

We now turn to consider the glassy disorder due to the dispersion of the $w_n$. 
Here we cannot trust the Born approximation because a small parameter is absent. 
However, without any approximation we can write the map in the form 
\be{10}
r_{n+1} \ \ = \ \ R_n \left( 1 - \frac{1}{r_n}\right) + 1 -\frac{\lambda}{w_n} 
\eeq
For $\lambda{=}0$ the zero momentum state is a solution as expected, 
irrespective of the disorder: the randomness in $R_n$ is not effective 
in destroying the ${r=1}$ fixed point. Therefore, for small $\lambda$, 
we can set without much error ${R_n=1}$. The formal argument that justifies 
this approximation is based on the linearization ${(r_{n+1}-1)=R_n(r_n-1)}$, 
and the observation that the product ${R_1R_2R_3...}$ remains of order unity. 
In \Fig{f2a} and \Fig{f2b} we verify numerically that setting ${R_n{=}1}$
does not affect the determination of~$\gamma$.

Having established that \Eq{e10} with ${R_n=1}$ in a valid approximation, 
we realize that it reduces to \Eq{e8}, with zero-average random term
\beq
f_n = -\lambda \left(\frac{1}{w_n}-\frac{1}{w_0} \right)
\eeq 
This random term corresponds to the diagonal-disorder of the standard Anderson model.  
Consequently, the implied definition of~$\epsilon$ via \Eq{e81} justifies 
the identification of the {\em harmonic} average~$w_0$ as the effective coupling.

We observe that there is an {\em emergent} small parameter, 
namely, the dispersion of~$f$, which is proportional to~$\lambda$ irrespective of the glassiness. 
Thus we have deduced a {\em duality} between ``strong" glassy off-diagonal disorder 
and the ``weak" diagonal disorder.
In the context of the dual problem we can use the Born approximation \Eq{e9} with 
\be{1211}
\mbox{Var}(f) \ \ = \ \ 
\left[ 2\sin\left(\frac{k}{2}\right) \right]^2  w_0^2 \,\mbox{Var}\left(\frac{1}{w}\right)
\eeq
leading to \Eq{FGRw} but with two important modifications 
with respect to FGR-based derivation: 
{\bf (i)} we realize that $w_0$ should be the harmonic average, as conjectured in the introduction; 
{\bf (ii)} we realize that the dispersion for off-diagonal disorder should be re-defined as follows:  
\be{12}
\sigma_{\perp}^2 \ \ := \ \ w_0^4 \ \mbox{Var}\left(\frac{1}{w}\right)
\eeq
For log-box distribution the FGR definition $\sigma_{\perp}^2 = \mbox{Var}(w)$ 
and the revised definition \Eq{e12} provide exactly the same result. 
But for {\em random distance} disorder the two prescriptions differ enormously. 
This is demonstrate in \Fig{f2a} and \Fig{f2b}, 
were we present our numerical results 
together with the theoretical predictions. 

Having adopted the revised definition \Eq{e12}, we still see in \Fig{f2a} and \Fig{f2b}
that the inverse localization length $\gamma$ is over-estimated as $k$ becomes larger. 
We can trace the origin of this discrepancy to the Lifshitz anomaly in the Anderson model.
The condition ${|\epsilon-\epsilon_0|<\epsilon_c}$ translates into ${\lambda > w_0^{-3}[\mbox{Var}(1/w)]^{-2}}$.
Thus the anomaly develops not at the band floor but as we go up in~$\omega$, 
where the inverse localization length becomes $\gamma \propto \omega^{4/3}$ 
instead of $\gamma \propto \omega^2$. 
To verify that this is indeed the explanation for the deviation 
we base our calculation on \Eq{e150}, namely
\be{151}
\gamma \approx \left( \frac{1}{2} \left( \frac{\sigma_{\perp}}{w_0} \right)^2 k^4 \right)^{\frac{1}{3}} 
\mathcal{K} \left[ \left( 2 \left( \frac{\sigma_{\perp}}{w_0} \right)^4 k^2 \right)^{-\frac{1}{3}}  \right]
\ \ \ \
\eeq
The anomaly appears whenever the argument of $\mathcal{K}(E)$ is small, 
meaning large $k$ rather than small $k$. 
The validity of this formula is numerically established 
in \Fig{f2a} and \Fig{f2b} with no fitting parameters.
We note that a slightly better version 
of \Eq{e151} can be obtained by replacing the~$k$s 
by appropriate trigonometric functions as implied 
by the remark after \Eq{e150} and \Eq{e1211}. 
But the numerical accuracy  is barely affected by such an improvement.


\begin{figure}

\includegraphics[clip,width=0.9\hsize]{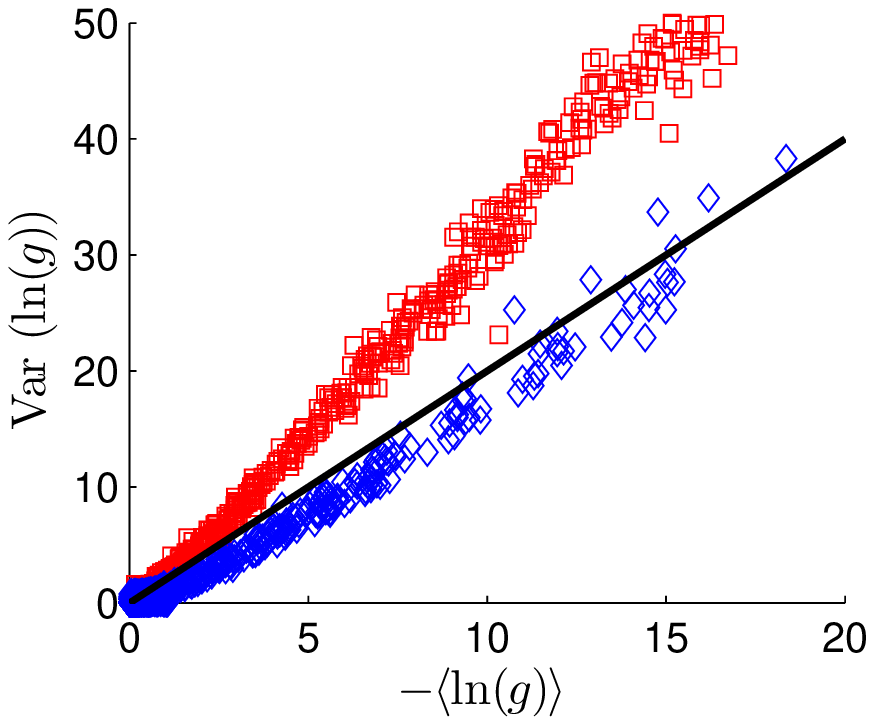}

\includegraphics[clip,width=0.9\hsize]{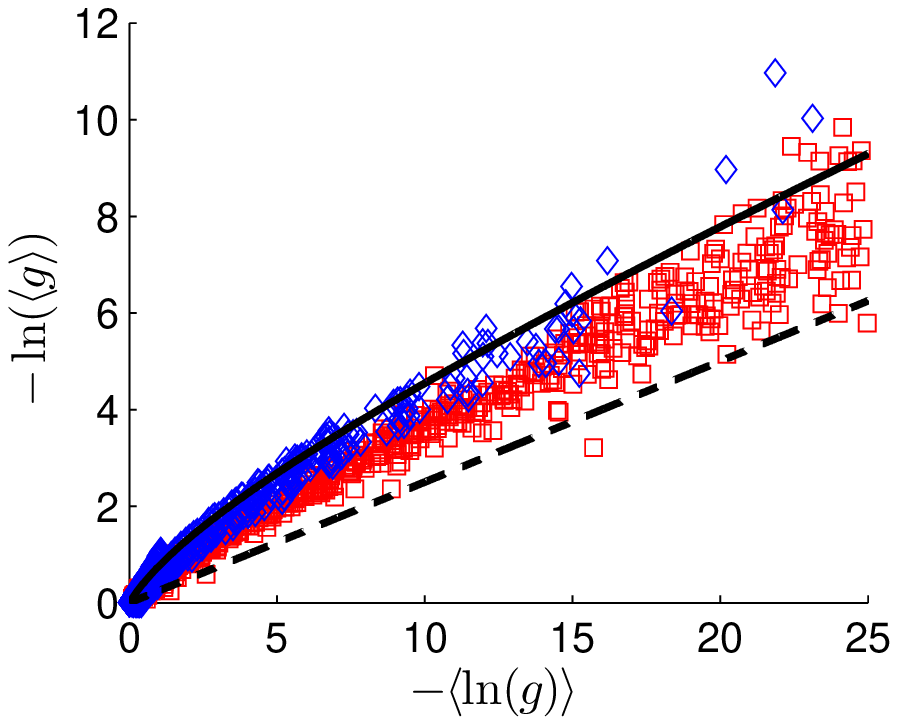}

\caption{
Testing one parameter scaling for glassy disorder. 
The variance of $\ln(g)$ (upper panel) and the log of the average $-\ln\langle g \rangle$ (lower panel)
are plotted against the scaling parameter $-\langle \ln(g) \rangle$. 
The calculation is done for the {\em random distance} disorder of \Eq{e3003} 
with ${s=15}$ (blue diamonds) and ${s=1.2}$ (red squares).
The solid line is the standard one parameter scaling prediction for weak disorder, 
and the dashed line is its asymptotic approximation. 
One observes that an anomaly develops as the disorder becomes glassy.}

\label{f3}
\end{figure}

\section{The average transmission}

For the calculation of the heat transport we have to know 
what is  ${ g(\omega) \equiv \langle g \rangle_{\omega} }$. 
Given $\gamma$ the common approximation 
is ${ \langle g \rangle \approx \eexp{-(1/2)\gamma L}}$.
But in-fact this asymptotic approximation can be trusted 
only for very long samples. More generally, assuming weak disorder, 
the following result can be derived~\cite{Abrikosov,Shapiro,Izrailev}:
\be{14}
\langle g \rangle \ = \ 
\int_0^\infty\mathrm{d} u
\ \frac{2\pi u\tanh(\pi u)}{\cosh(\pi u)}
\ \mathrm{e}^{-\left[\left(\frac{1}{4}+u^2\right)\gamma L\right]}
\label{gabrik}
\eeq
The question arises whether this relation can be trusted 
also in the case of a glassy disorder, where the 
one-parameter scaling assumption cannot be justified.
This is tested in \Fig{f3}. For weak disorder (large $s$) 
the expected relation between the first and second moments 
of $x=-\ln(g)$ is confirmed, namely ${\text{Var}(x)=2\langle x \rangle}$.
For strong glassy disorder (small $s$) clear deviation 
from this relation is observed. 

Still we see in \Fig{f3} (lower panel) that the failure of one-parameter scaling 
is not alarming for $\langle g \rangle = \langle \exp[-x] \rangle$.
The exact calculation of the integral is the solid line,
while the asymptotic result $\exp[-(1/2)\langle x \rangle]$ 
is indicated by dashed line. Note that the latter 
implies ${\langle g \rangle = \exp[(1/4)\langle\ln(g)\rangle]}$. 
We realize that the asymptotic approximation might be poor, 
but the exact calculation using \Eq{e14} is quite satisfactory.

\begin{figure}

\includegraphics[clip,width=0.9\hsize]{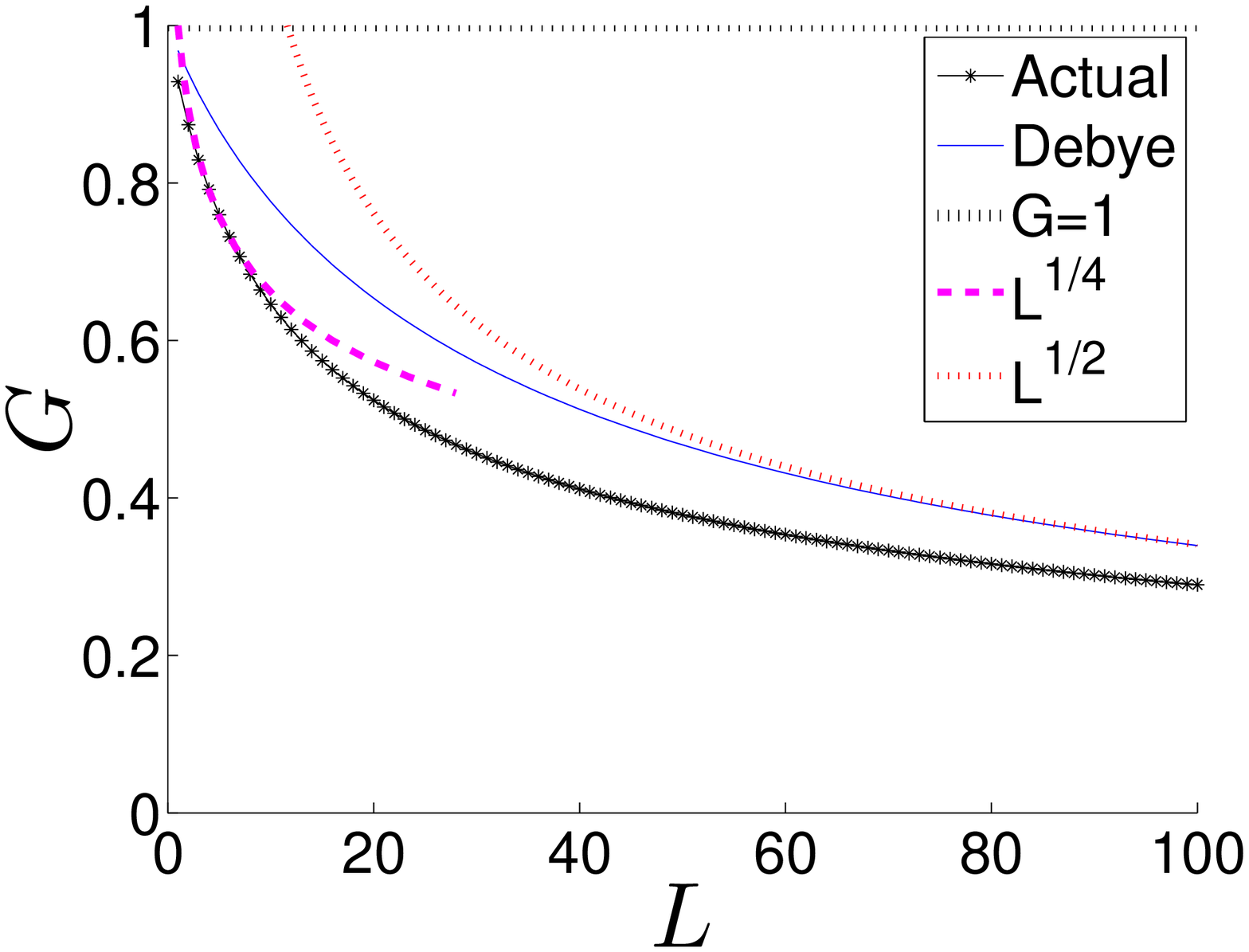}
\includegraphics[clip,width=0.9\hsize]{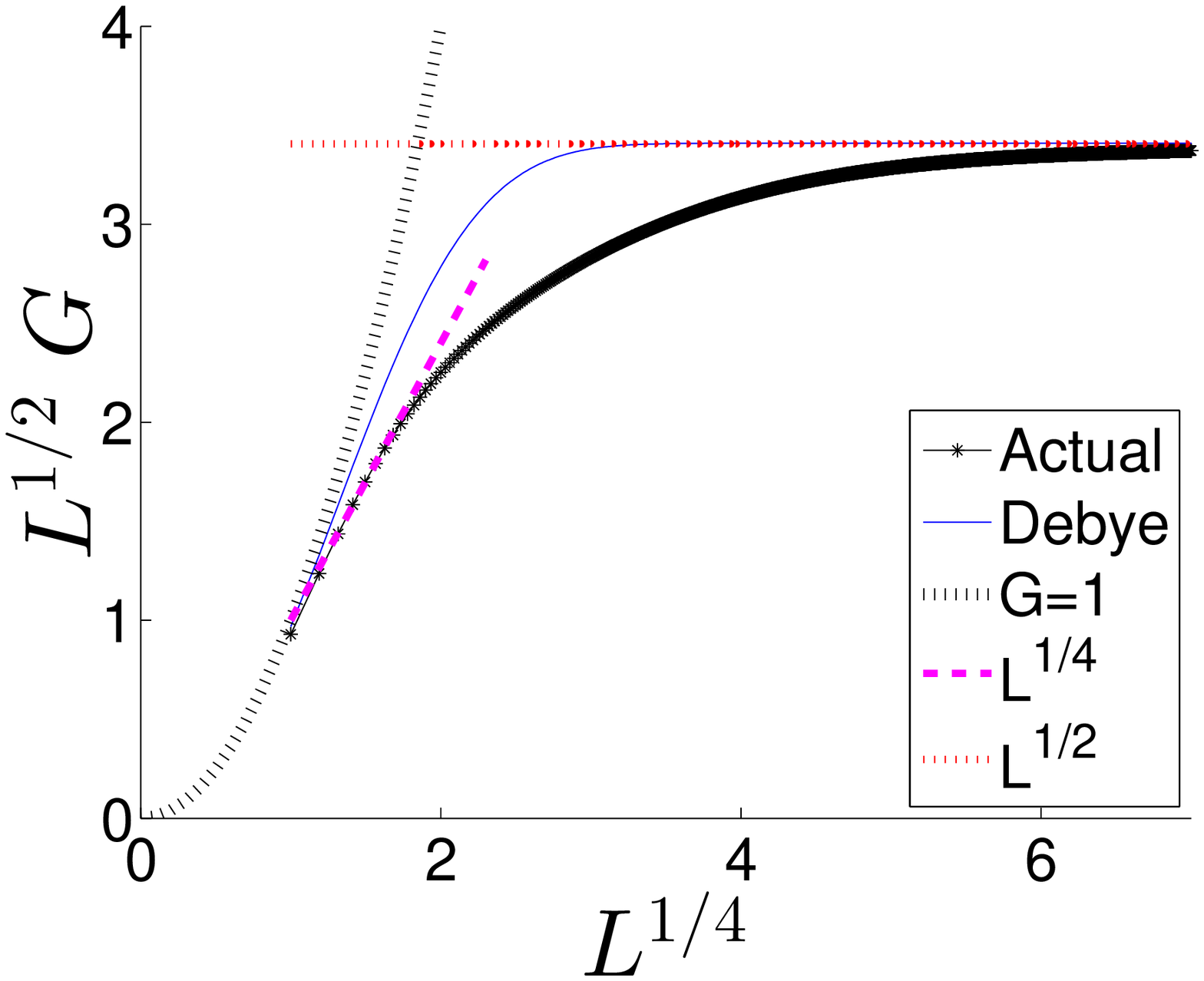}

\caption{
The heat conductance $G$ as a function of $L$ is calculated using \Eq{e17} with \Eq{e14}.
For the black solid line with symbols we use $\gamma(k)$ that is based on the numerical
results that have been obtained for a {\em random barrier} disorder as in \Fig{f2a} with ${\sigma=1}$.  
For the blue solid line we use the Debye approximation for the density 
of states, and extrapolate the initial $\gamma\propto k^2$ dependence up to the cutoff $k=\pi$.  
On the lower panel we plot $\sqrt{L}G$ as a function of $L^{1/4}$ 
in order to highlight the ${\alpha=1/2}$ (dotted red) and the ${\alpha=1/4}$ (dashed red)
asymptotic dependence for long and short samples respectively. 
The black dotted line is $G=1$. }

\label{f4}
\end{figure}

\section{Heat conductance}

Following  \cite{D08,D01,DL08} the expression for the rate of heat flow from 
a lead that has temperature $T_H$ to a lead that has temperature $T_C$ is 
\beq
\dot{Q} \ = \ 
\frac{T_H{-}T_C}{2} \int_{0}^{\infty} \frac{d\omega}{\pi} \ \mathcal{T}(\omega) 
\ \equiv \ G \, (T_H{-}T_C) \ \ 
\label{qg}
\eeq
Here $\mathcal{T}(\omega)$ is a complicated expression that reflects 
the transmission of the sample.  
If we were dealing with incoherent or non-linear transport \cite{dubi}, 
it would be possible to justify the Ohmic expression ${\mathcal{T}(\omega)=\ell/L}$, 
where $\ell$ is the inelastic mean free path. 
But we are dealing with an isolated harmonic chain, 
therefore ${\mathcal{T}(\omega)}$ is determined by 
the couplings of the eigenmodes to the heat reservoirs
at the left and right leads. 
In analogy to mesoscopics studies \cite{D08,D01} one can argue that
\beq
\mathcal{T}(\omega) \ \ \approx \ \ g(\omega) \ \mathcal{T}^{(0)}(\omega)
\eeq
where $\mathcal{T}^{(0)}(\omega)$ refers to a non-disordered sample, 
and $g(\omega)$ is the disordered averaged transmission. 
For ``fixed boundary conditions" ${\mathcal{T}^{(0)}(\omega)\sim \eta_0^2\omega^2}$,
where the damping rate $\eta_0$ characterizes the contact point.
In contrast, for ``free boundary condition" one obtains ${\mathcal{T}^{(0)}(\omega)\approx 1}$, 
which is the most optimal possibility. In the latter case  
\be{17}
G \ \ = \ \ \frac{c}{2} \int_0^{\pi} \frac{d k}{\pi} \ g(\omega_k) 
\eeq
The standard approach is to use two {\em incompatible} approximations:
On the one hand one use the asymptotic estimate ${g(\omega) \sim \eexp{-(1/2)\gamma(\omega) L}}$ 
which holds for {\em long} samples for which $\gamma L\gg 1$. 
On the other hand one extends the upper limit of the integration to infinity, 
arguing that the major contribution to the integral comes from small $\omega$ values.
In the absence of pinning $\gamma(\omega)\propto\omega^2$, hence by rescaling 
of the dummy integration variable it follows that the result of the integral 
is precisely $\propto 1/\sqrt{L}$. 
We shall discuss in the next paragraph the limitations of this prediction. 
Going on with the same logic we can ask what happens in the presence of 
a weak pinning potential. Using a saddle-point estimate 
we get 
\beq
G \ \ \sim \ \  \frac{1}{\sqrt{L}} \ \exp\left[ -(1/2) \gamma_0 L \right]
\eeq
where $\gamma_0$ is the minimal value of $\gamma(\omega)$. 
From \Eq{FGRw} with \Eq{FGRv} we deduce $\gamma_{0} \propto b^{-\eta}$, with $\eta=7/2$.
This explains the leading exponential dependence on the length 
that has been observed in \cite{BZFK13}. 
However the above calculation fails in explaining the 
sub-leading $L$~dependence that survives in the absence of pinning.
Namely it has been observed that instead of $1/L^{\alpha}$ with $\alpha=1/2$
the numerical results are characterized by the super-optimal value ${\alpha \approx 1/4}$.

\section{Beyond the asymptotic estimate}

We now focus on the $L$ dependence that survives in the absence of a pinning potential.
As already note that deviation from the $1/\sqrt{L}$ law is related to two incompatible 
approximations regarding the~$\gamma$ dependence of~$g$ 
and the upper limit of of the integration in \Eq{e17}.    
We can of course do better by using the analytical expression \Eq{e14}.
For the density of states we can use either numerical results 
or optionally we can use the Debye approximation. The latter 
may affect the results {\em quantitatively} but not qualitatively.
Within the framework of the Debye approximation we assume 
idealized dependence $\gamma(\omega_k) \propto \sigma_{\perp}^2 k^2$ 
in accordance with \Eq{FGRw}, up to the cutoff at $k=\pi$.
The result of the calculation is presented in \Fig{f4}.  
On the lower panel there we plot $\sqrt{L}G$ as a function of $L^{1/4}$ 
in order to highlight the ${\alpha=1/2}$ and the ${\alpha=1/4}$ 
asymptotic dependence for long and short samples respectively. 
Note that in the latter case, as in \cite{BZFK13},  
a small offset has been included in the fitting procedure.
We do not think that the ${\alpha=1/4}$ dependence is ``fundamental". 
The important message here is that a straightforward application 
of an analytical approach can explain the failure of the $1/\sqrt{L}$ law. 
We also see that the numerical prefactor of the $1/L^{\alpha}$ dependence 
is sensitive to the line-shape of the large~$k$ cutoff, 
hence it is not the same for the numerical spectrum and for its Debye approximation.

\section{Conclusions} 

We have considered in this work the problem 
of heat conduction of quasi one-dimensional ($b\gg1$) 
as well as single channel harmonic chains; 
addressing the effects of both glassy disorder (couplings) and pinning (diagonal disorder).
We were able to provide a theory for the asymptotic exponential 
length ($L$) and bandwidth ($b$) dependence; as well as for the 
algebraic $L$~dependence in the absence of a pinning potential. 
A major observation along the way was the duality between 
glassy disorder and weak Anderson disorder. That helped 
us to figure out what is the effective hopping~$w_0$, 
hence establishing a relation to percolation theory.
It also helped us to go beyond the naive Born approximation
that cannot be justified for glassy disorder, 
and to identify a (dual) Lifshitz anomaly in 
the $\omega$ dependence of the transmission. 
Finally, we have established that known results from 
one-parameter scaling theory can be utilized in order 
to derive the non-asymptotic $L$ dependence of the heat conductance.


\ \\
{\bf Note added after acceptance.-- }
It has been observed in \cite{Chalker} that the one-dimensional localization problem
with the distribution \Eq{e3003} describes in a universal way    
the phononic excitations of a one-dimensional Bose-Einstein condensate in a random potential: 
the superfluid regime is percolating (${s{>}1}$), while the Mott-insulator regime
corresponds to the ``$s{=}0$" random-barrier distribution \Eq{e3002}. The localization 
length in the anomalous regime ${1{<}s{<}2}$ had been worked out in~\cite{Ziman}, 
leading to ${\gamma \propto \omega^s}$, while ${\gamma \propto \omega^2}$ applies if ${s{>}2}$.

\ \\
{\bf Acknowledgements.-- }
We thank Boris Shapiro (Technion) for helpful comments. 
This research has been supported by  by the Israel Science Foundation (grant No. 29/11), 
and by the NSF Grant No. DMR-1306984.

\vspace*{-0.2cm}

\appendix 

\section{Localization in white noise potential}

There is an analytical expression for the counting function of the energy-levels 
for a particle in a white-noise disordered potential.
We cite Eq(1.62) of \cite{halperin}:
\be{111}
\mathcal{N}(E) \ = \ \frac{1}{\pi^2}\Big([\text{Ai}(-2E)]^2 + [\text{Bi}(-2E)]^2 \Big)^{-1}
\ \ \ \ \ 
\eeq
In this expression the levels are counted per unit length;  
Ai and Bi are the Airy functions; and the energy $E$ is 
expressed in scaled units. It reduces to ${\mathcal{N}_0(E)=(2E)^{1/2}/\pi}$, 
as expected, in the limit of zero disorder.
Here we use \Eq{e111} as an approximation that applies  
at the bottom of the band, where the scaled energy is  
\beq
E \ \ = \ \  \left[ 2\frac{[\text{Var}(v_n)]^2}{w_0} \right]^{-1/3} \, \lambda  
\eeq

\begin{figure}

\includegraphics[width=0.9\hsize]{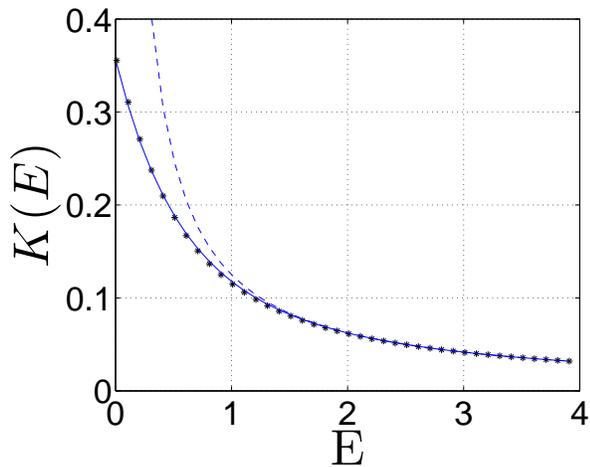}

\caption{
The function $\mathcal{K}(E)$ calculated numerically (symbols). 
The solid and dashed lines are our approximation \Eq{e182} 
and the asymptotic $1/(8E)$ respectively.    
}

\label{fKE}
\end{figure}

The inverse localization length is related to 
the counting function by the Thouless relation, 
namely Eq(19) of \cite{thouless}.
Note that the subsequent formulas there are 
confused as far as units are concerned. 
In order to remedy this confusion we define 
\beq
\mathcal{K}(E) \ \ = \ \ \int \frac{\mathcal{N}(z)-\mathcal{N}_0(z)}{E-z} dz
\eeq
and write the Thouless relation as follows:
\beq
\gamma \ = \ \left( \frac{\text{Var}(v_n)}{2w_0^2} \right)^{1/3} 
\mathcal{K}\left[ \left( 2\frac{[\text{Var}(v_n)]^2}{w_0} \right)^{-1/3} \, \lambda  \right]
\ \ \ \ \ \ \ \ \ 
\eeq
For large $E$ one obtains ${\mathcal{K}(E)\approx 1/(8E)}$
and the Born approximation is recovered:
\beq
\gamma \ \ = \ \ \frac{\text{Var}(v_n)}{8w_0 \, \lambda} 
\ \ = \ \ \frac{\text{Var}(v_n)}{8w_0^2 \, k^2} 
\eeq
For practical purpose we find that an excellent approximation for any ${E>0}$ is 
provided by 
\be{182}
\mathcal{K}(E) \ \ \approx \ \ \left[1-\exp\left(-\frac{E}{E_0}\right)\right] \frac{1}{8 \, E}
\eeq 
where $E_0=0.35$. See \Fig{fKE} for demonstration.


\clearpage

\begin{thebibliography}{99}
\bibitem[a]{rmrkA}
Given an eigenstate $\psi$ we define $p_n=|\psi_n|^2$.
The normalization is such that $\sum p_n=1$. 
The participation number PN$=[\sum p_n^2]^{-1}$ 
is a measure for the number of sites that are occupied 
by the eigenstate.
\end{thebibliography}
\end{document}